\documentclass[showpacs,superscriptaddress,groupedaddress]{revtex4-1}  
\usepackage{graphicx}  
\usepackage{bm}        
\usepackage{amssymb}   
\usepackage{amsmath}
\usepackage{ mathrsfs }
\usepackage{pdfpages}
\def\mathclap#1{\text{\hbox to 0pt{\hss$\mathsurround=0pt#1$\hss}}}
\usepackage{float}

\begin{document}

\widetext

\title{The Shape of a Stretched Polymer}
%
%
\author{Alberto S. Sassi} \affiliation{Laboratoire de Biophysique Statistique, Ecole Polytechnique F\' ed\' erale de Lausanne (EPFL), CH-1015 Lausanne, Switzerland}
\author{Salvatore Assenza} \affiliation{Laboratoire de Biophysique Statistique, Ecole Polytechnique F\' ed\' erale de Lausanne (EPFL), CH-1015 Lausanne, Switzerland}
\author{Paolo De Los Rios} \affiliation{Laboratoire de Biophysique Statistique, Ecole Polytechnique F\' ed\' erale de Lausanne (EPFL), CH-1015 Lausanne, Switzerland}
%
%
%
\vskip 0.25cm
\begin{abstract}
The shape of a polymer plays an important role in determining its interactions with other molecules and with the environment, and is in turn 
affected by both of them. As a consequence, in the literature the shape properties of a chain in many different conditions have been investigated.
Here, we characterize the shape and orientational properties of a polymer chain under tension, a physical condition typically realized both in 
single-molecule experiments and \textit{in vivo}. By means of analytical calculations and Monte Carlo simulations, 
we develop a theoretical framework which quantitatively describes these
properties, highlighting the interplay between external force and chain size in determining
the spatial distribution of a stretched chain.

\end{abstract}

\pacs{}
\maketitle

Polymer chains are 
strongly-fluctuating objects.
Due to their soft nature, they do not have well-defined shapes, but rather adapt
their conformational ensembles to the environment surrounding them. 
At the same time, the shape of a polymer affects the way it interacts with other
molecules in solution, \textit{e.g.} quantitatively determining the excluded volume of the chain,
thus ultimately influencing the thermodynamic properties of the solution \cite{rubinstein, minton}.
In a poor solvent, 
polymers collapse into a roughly spherical shape, as seen for example in the case of 
single-chain globular proteins \cite{dima_thirumalai_globular_protein}. 
In the case of $\theta$- or good-solvent conditions,
conformations can fluctuate more wildly and the shape of a polymer is more sensitive to 
the environment. In the last two decades, many theoretical and experimental works have 
investigated how the shape depends on 
several factors such as 
confinement \cite{thirumalai_confinement, dekker_confinement, micheletti_confinement},
topology \cite{preusser_topology, frey_topology, millett_topology} or crowding 
\cite{dima_thirumalai_globular_protein, denton_crowding, thirumalai_crowding, stafshede_crowding}, which
are relevant to many cellular processes.

\textit{In vivo} biopolymers are often under the action of mechanical forces. For example,
during replication DNA is repeatedly pulled and twisted by enzymes \cite{bustamante_review},
and proteins are actively pulled by chaperones across membranes and out of 
ribosomes \cite{matlack, neupert, theg, qian, bustamante2015,frontiers}.
Moreover, the recent development of several single-molecule techniques such as AFM, Optical and Magnetic 
Tweezers has stimulated a large amount of experimental works involving polymer stretching \cite{RitortReview}.
Surprisingly, in spite of the great importance of stretched chains for both \textit{in vivo} and single-molecule studies,
a proper account for the effect of an external pulling force on the shape of a polymer has
been lacking, even at the simplest level. Here we compute, by 
means of both analytical calculations and Monte Carlo (MC) simulations, 
some key global quantities characterizing the shape of a stretched chain. Particularly,
we focus on the simple case of the Freely Jointed Chain (FJC) model, 
a useful representation of a polymer in a good solvent where 
the conformation of a chain made by $N+1$ monomers is described 
as a three-dimensional Random Walk $\{\bm{R}_i\}_{i=0}^N$, where the segments
$\bm{r}_i\equiv\bm{R}_{i}-\bm{R}_{i-1}$ have all 
equal length $b$ \cite{rubinstein}.

We first address the exact computation of the components of the end to end vector $\bm{R_e}\equiv\bm{R}_N-\bm{R}_0=\sum_{i=1}^N \bm{r}_i$ of a stretched FJC and the relative variances. 
At equilibrium, the conformations of the chain are distributed according 
to the stretching energy $-\bm{f}\cdot\bm{R_e}$, where $\bm{f}$ is the applied external force. The key point of the derivation is that the force is coupled independently to each segment: $-\bm{f}\cdot\bm{R_e}=-\sum_i\bm{f}\cdot\bm{r_i}$. It is thus possible to calculate the partition function of a single segment, that is $q(\beta fb)=\int d\bm{r}\exp(\beta\bm{f}\cdot\bm{r_i})\delta\left(\left|\bm{r_i}\right|-b\right)=4\pi b^2\sinh(\beta fb)/\beta fb$ and then compute the partition function of the whole chain $Q$ as $Q=q^N$.  
Starting from $Q(\beta fb)$ and choosing a reference frame with the $z$ axis oriented along the force,
a straightforward computation leads to the well-known result \cite{rubinstein}
$\left<\bm{R_e}\right>=Nb\,\mathcal{L}\,\bf{e_z}$, where
$\gamma\equiv \beta fb$ and
$\mathcal{L}\equiv\coth(\gamma)-1/\gamma$ is the Langevin function.
Analogously, the variances of the three components of $\bm{R_e}$ are
$\sigma^2_x=\sigma^2_y=Nb^2g$ and $\sigma^2_z=Nb^2a$, where $g\equiv \mathcal{L}/\gamma$ 
and $a\equiv1-2g-\mathcal{L}^2$.
Collecting these results, we obtain
the mean value of the square of the end to end distance:
\begin{equation}
\left<\bm{R_e}^2\right>=Nb^2(1-\mathcal{L}^2)+N^2b^2\mathcal{L}^2 \ ,
\end{equation}
and of the radius of gyration (see section S1)
\begin{equation}
 \left<R_g^2\right>=\frac{1}{6}Nb^2+\frac{1}{12}N^2b^2\mathcal{L}^2\,.
 \label{eqn:rg}
\end{equation}
As expected, we find the unperturbed results 
$\left<\bm{R_e}^2\right>\simeq Nb^2, \left<R_g^2\right>\simeq Nb^2/6$ when $\gamma\rightarrow 0$, and the typical values 
of a rod $\left<\bm{R_e}^2\right>\simeq N^2b^2, \left<R_g^2\right>\simeq N^2b^2/12$ in the limit of infinite forces. The transition between the two regimes takes place at a force $\gamma_c \sim 1/\sqrt{N}$, in
agreement with previous perturbative results \cite{Neumann1}.

The radius of gyration is a useful quantity to measure the spatial extent of a macromolecule. 
Nevertheless, it cannot capture the presence of anisotropies, which ultimately 
determine the overall shape of the polymer.
In this respect, a more suitable tool is the inertia tensor
$\mathcal{T}$, whose elements are
$\mathcal{T}_{\alpha\beta} = \frac{1}{N+1}\sum_{i=0}^{N}\left(\alpha_{i}-\alpha_{cm}\right)\left(\beta_{i}-\beta_{cm}\right)$,
where $\alpha,\beta\in\{x,y,z\}$ and $\bm{R}_{cm}\equiv(x_{cm},y_{cm},z_{cm})$ is the position of the center of mass of the chain \cite{rudnickart}. 
The eigenvalues $\lambda_1\ge\lambda_2\ge\lambda_3\ge0$ of $\mathcal{T}$ are proportional 
to the square of the semiaxes of the ellipsoid which best approximates the shape of the polymer, while the corresponding 
eigenvectors describe its orientation. Moreover, by construction $R_g^2=\lambda_1+\lambda_2+\lambda_3$.
Strikingly, when it comes to shape properties even the simplest models for polymers give non-trivial results. 
As already noticed in an early work by Kuhn \cite{kuhn}, in spite of the spherical symmetry of the system,
the typical shape of an unperturbed FJC is an elongated ellipsoid, whose random orientation 
preserves the overall symmetry.
Numerical computations have shown that the ensemble averages of the eigenvalues 
of an unperturbed FJC are in the ratios $\left<\lambda_1\right>:\left<\lambda_2\right>:\left<\lambda_3\right>=11.8:2.7:1$,
thus outlining a pronounced asymmetry in the spatial distribution of the monomers \cite{rudnickart}. 

\begin{figure}[!h]
 \includegraphics[scale=0.21]{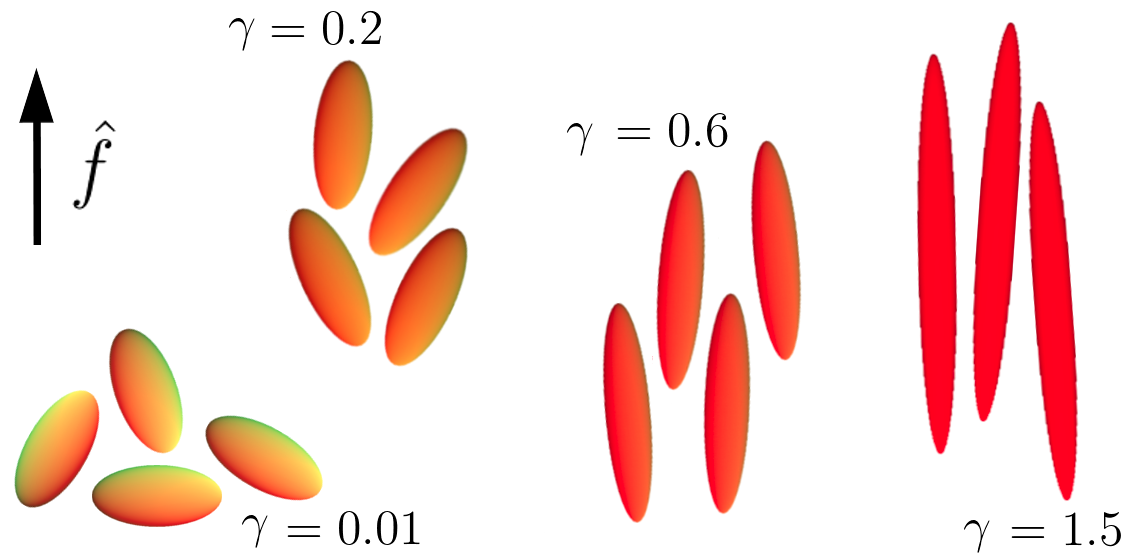}
 \caption{\label{fig:orientation_scheme} Schematic illustration of the effect of an external force
on a FJC. The arrow indicates the direction of the external force.
All the ellipsoids shown have axes ratios corresponding to the average 
values of the eigenvalues for a chain made of $N=200$ segments and for the values
of $\gamma$ reported in the figure.
Particularly, we show the projection corresponding to the two largest eigenvalues.
To ease visualization, the absolute size of the ellipsoids was chosen arbitrarily, thus relative 
sizes of ellipsoids drawn for different values of $\gamma$ do not reflect the real values.}
\end{figure}
\begin{figure}[!h]
 \includegraphics[scale=0.24]{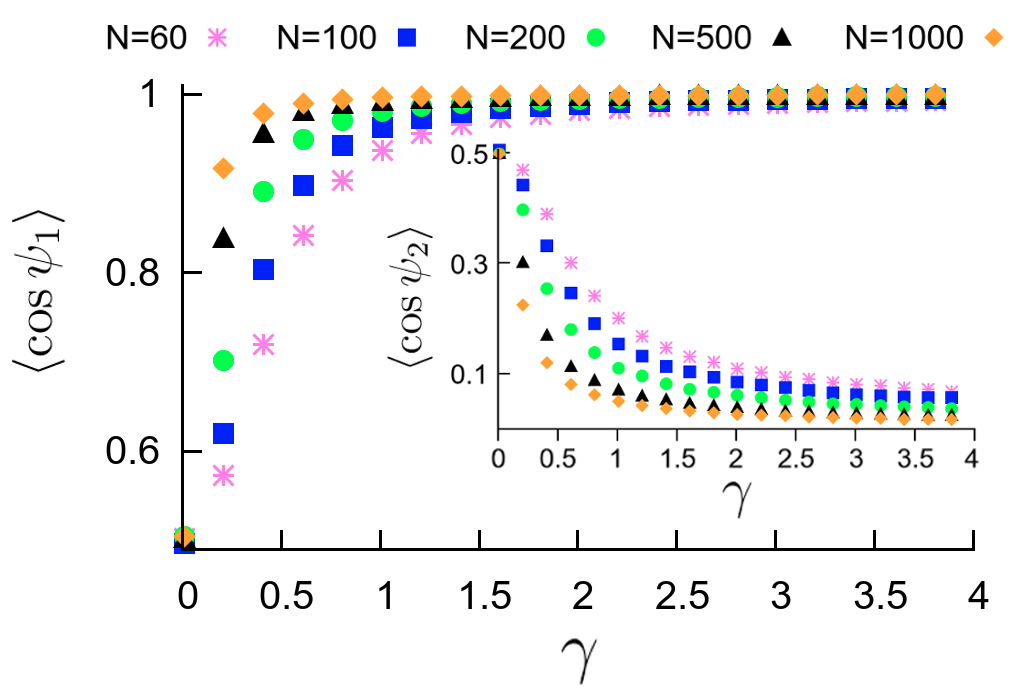}
 \caption{\label{fig:angles} Orientation of the ellipsoid enveloping a stretched FJC
as a function of the adimensionalized force $\gamma$. 
In the plot we report for several values of $N$
the average cosine of the angles $\psi_1$ and $\psi_2$ (inset) between the external force 
 and the eigenvectors corresponding to $\lambda_1,\lambda_2$ respectively.}
\end{figure}
\begin{figure}[!h]
 \includegraphics[scale=0.23]{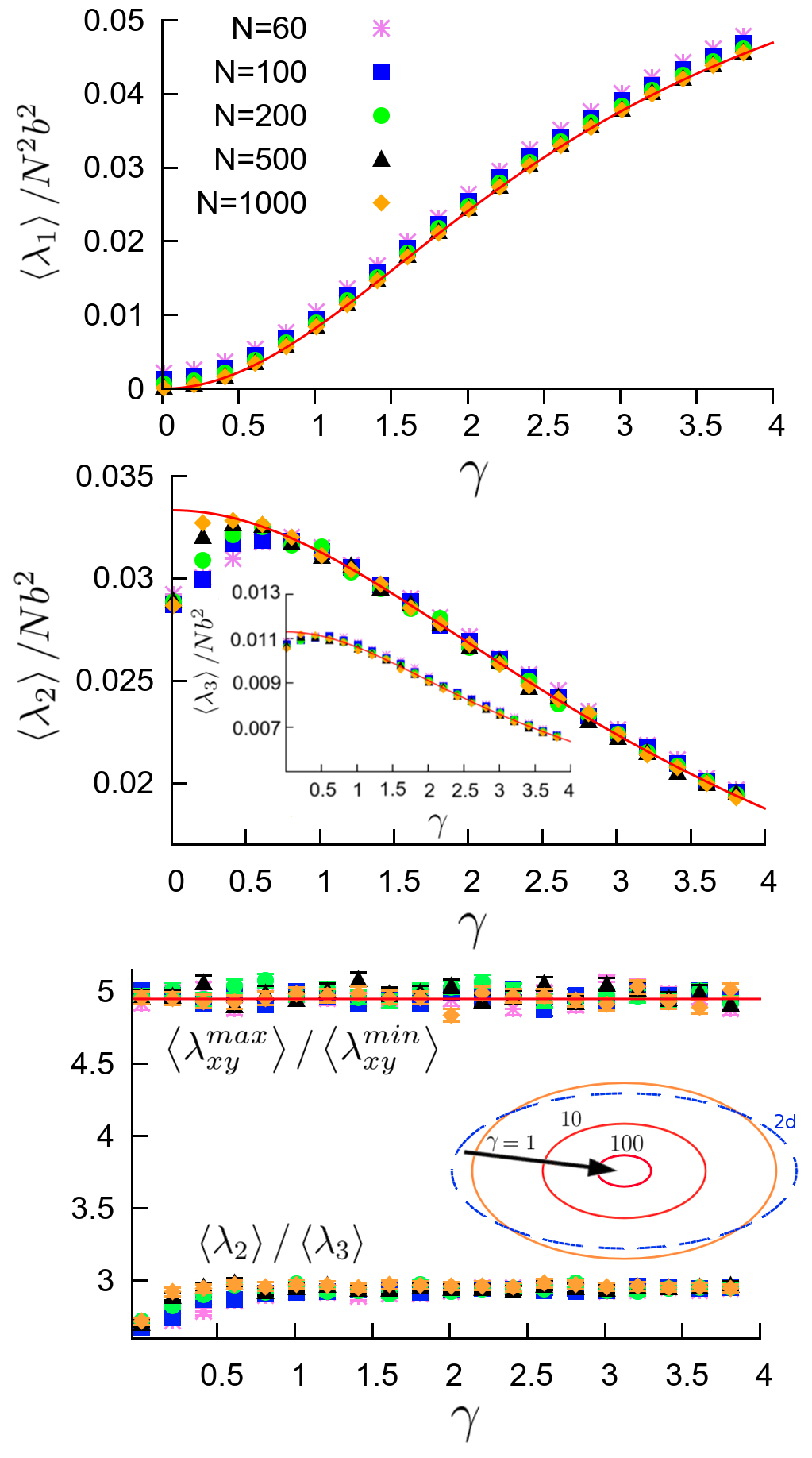}
 \caption{\label{fig:lambdas} Top and center panels: normalized eigenvalues $\lambda_1/N^2b^2$ (top), $\lambda_2/Nb^2$ (center)
 and $\lambda_3/Nb^2$ (center inset) as a function of the adimensionalized force $\gamma$, for several 
 values of chain size $N$. The continuous curves are the corresponding fitting functions reported in
 the main text. Bottom panel: aspect ratio of the $xy$ projection of the chain and 
 of the transverse section of the ellipsoid. The continuous red line shows the aspect ratio of a pure two-dimensional
 FJC, which we computed by means of MC simulations and found to be equal to $4.95\pm 0.01$. Inset:
 sketch of the shape of the transverse section of the chain
for $\gamma=1,10,100$, where the axes are proportional to $\sqrt{\lambda_2}$
and $\sqrt{\lambda_3}$ (continuous lines). The arrow indicates the direction of increasing $\gamma$. The 
blue dashed line sketches the typical shape of a two-dimensional FJC.}
\end{figure}
In order to characterize the shape of a stretched chain, we 
evaluated the eigenvalues of the inertia tensor and the corresponding eigenvectors
by means of Monte Carlo simulations (for details see section S2 in the Supplemental Material). 
In Fig.\ref{fig:orientation_scheme} we show a schematic picture recapitulating 
the evolution of the polymer as the tension is increased.
As the picture qualitatively shows, the effect of an external force on a chain is twofold.
For low values of $\gamma$, the tension mostly affects the orientation of the 
enveloping ellipsoid, aligning it along its direction \cite{orlandinirev}
(left region in Fig.\ref{fig:orientation_scheme}).
Correspondingly, in Fig.\ref{fig:angles} and in Fig.S1 in the Supplemental Material
we show the average value of $\cos\psi_s$, 
where $\psi_s$ is the angle between the applied force and the eigenvector 
corresponding to the eigenvalue $\lambda_s$. Starting from the typical value of random orientations ($1/2$), the cosine rapidly approaches $1$ in the case of $\lambda_1$ (Fig.\ref{fig:angles})
and $0$ for $\lambda_2$ (Fig.\ref{fig:angles} inset) and $\lambda_3$ (Fig.S1),
which correspond to an ellipsoid with the principal axis oriented along the force. 
It is worth noting that the cosine approaches its large-$\gamma$ value more rapidly for larger sizes of the chain. After this ``dipole-like'' regime,
the tension strongly deforms the polymer, thus leading 
to an increase in the anisotropy of its shape (right region in Fig.\ref{fig:orientation_scheme}).
However, we note that, due to the monotonicity of $\mathcal{L}^2$, $R_g^2$ increases monotonically
from its unperturbed value to the rodlike limit, thus a weak streching 
is present also at low forces.
After the ellipsoid has aligned to the force,
$\lambda_1$ gives the largest contribution to
$R_g^2$. From Eq.(\ref{eqn:rg}), we thus expect $\left<\lambda_1\right>\simeq N^2 b^2 L^2/12$,
which is verified by MC simulations (Fig.\ref{fig:lambdas} top),
with small but systematic deviations that decrease for longer chains.
For large forces, the two remaining eigenvalues
are expected to behave as the transverse contributions to $R_g^2$, and
are governed by the fluctuations of the chain on the plane perpendicular to the force. 
Comparing the formula for $\sigma_{x,y}$, we thus predict that in this regime
$\left<\lambda_{2}\right>\simeq c_2Nb^2g$ and $\left<\lambda_{3}\right>\simeq c_3Nb^2g$,
where $c_2$ and $c_3$ are numeric constants.
This ansatz is confirmed in Fig.\ref{fig:lambdas} center,
where the continuos curves are obtained by tuning the coefficients in order to globally fit the MC data obtained for $\gamma>1$ and correspond to $c_2= 0.100 \pm 0.001$ and $c_3= 0.034\pm 0.001$. 
Intriguingly, $\left<\lambda_2\right>$ and $\left<\lambda_3\right>$ show a non-monotonic
behavior for small forces. More in detail, starting from $\gamma=0$ they increase 
up to a maximum, after which they decrease according to the large-$\gamma$ behavior. A comparison
with the corresponding values of $\left<\cos\psi_2\right>,\left<\cos\psi_3\right>$ (Fig.\ref{fig:angles} and Fig.S1) shows that the 
range of forces with increasing $\lambda_2,\lambda_3$ corresponds to a regime where the ellipsoid has still to align with the force. Therefore, an intuitive explanation of this phenomenon is that, due to the 
random orientation of the polymer (see left region in Fig.\ref{fig:orientation_scheme}),
on average in this regime the force deforms the ellipsoid 
almost isotropically, thus leading to an 
increase of all the eigenvalues. In contrast, after a perfect alignment has been achieved
(right region in Fig.\ref{fig:orientation_scheme}), 
only $\left<\lambda_1\right>$ keeps growing, while the two smaller eigenvalues shrink due to the 
smaller and smaller fluctuations in the directions perpendicular to the force.
Notably, in the large-force regime $\left<\lambda_2\right>/\left<\lambda_3\right>=c_2/c_3\simeq 3$ 
for all values of $N$ (Fig.\ref{fig:lambdas} bottom), 
implying that the section of the ellipsoid shrinks while preserving 
a universal shape independent
of the size of the polymer (inset).
Since for $\gamma\gg 1$ the chain is almost aligned with the external force, one would expect
to identify the universal section of the ellipsoid with the projection of the chain onto the $xy$ plane that,
because of the independence of the three directions of a Random Walk, has the same features of
a two-dimensional FJC.  
The shape properties of the $xy$ projection of the chain 
are obtained by diagonalizing the submatrix of the inertia tensor identified by 
the elements 
$\mathcal{T}_{xx},\mathcal{T}_{xy},\mathcal{T}_{yy}$.
As we report in Fig.\ref{fig:lambdas} bottom, the ratio between the averages of its eigenvalues
$\lambda_{xy}^{max}$ and $\lambda_{xy}^{min}$ closely follows the behavior of a two-dimensional 
FJC (red continuous curve), but is larger than $\left<\lambda_2\right>/\left<\lambda_3\right>$. 
Why do the transverse section and the $xy$ projection of the 
ellipsoid have different shapes? Qualitatively, the key point is that, although the main axis of the ellipsoid
becomes more and more aligned with the external force, the value of 
$\lambda_1$ increases with $\gamma$, therefore its projection onto the $xy$ 
plane is comparable to the contributions coming from $\lambda_2$ and $\lambda_3$.
From a quantitative point of view, we note that the total contribution to $\left<R_g^2\right>$
involving the $x$ and $y$ coordinates is equal to $Nb^2g/3\simeq 0.33 Nb^2g$ (see section S1).
Nevertheless, the large-force formulas provided above for $\left<\lambda_2\right>,\left<\lambda_3\right>$
show that the sum of the smaller eigenvalues is 
$\left<\lambda_2\right>+\left<\lambda_3\right>=(c_2+c_3)Nb^2g\simeq 0.13Nb^2g$. Therefore, also 
$\left<\lambda_1\right>$ is expected to give a significant contribution to the $x,y$ projection of the FJC equal to 
$c_1Nb^2g$, where $c_1=1/3-c_2-c_3\simeq 0.20$ (as we show in section S3, the same result can be found starting
directly from the MC data for $\left<\lambda_1\right>$). These results show that the
isotropically-shrinking transverse section 
of the ellipsoid cannot be identified with the $xy$ projection of the FJC, thus outlining 
a novel universal behavior in the shape of polymers.

\begin{figure}[!h]
 \includegraphics[scale=0.21]{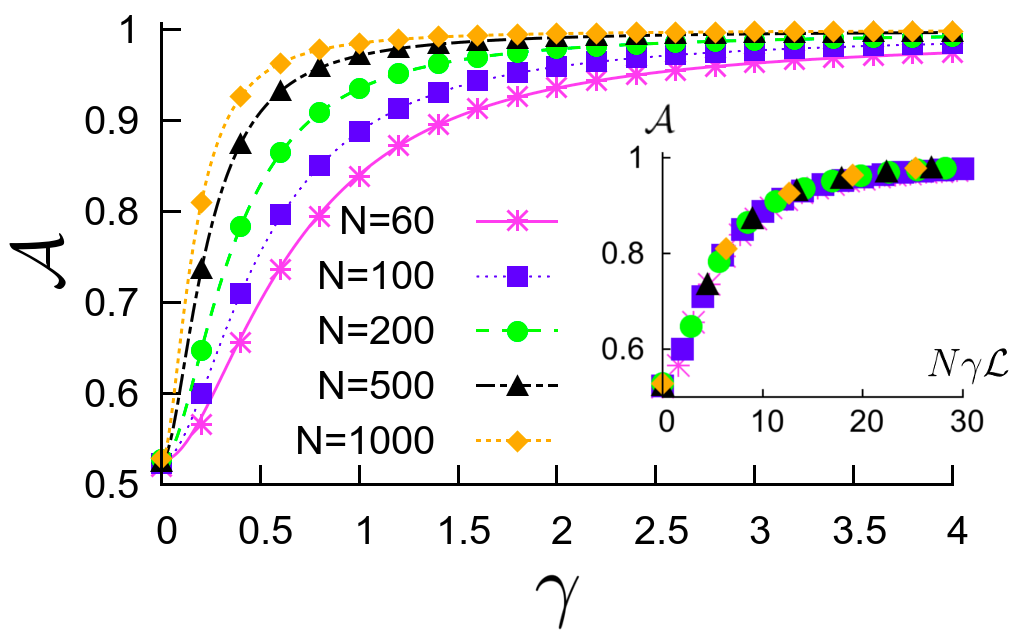}
 \caption{\label{fig:asph} Comparison between Monte Carlo data and exact formula (Eq. (\ref{eqn:asphericity_stretch}))
 for the aspericity $\mathcal{A}$
 as a function of the adimensionalized force $\gamma$, for several values of $N$. If the data are plotted as a function of $N\gamma\mathcal{L}$,
 all the sets approximately collapse onto a universal curve (inset).}
\end{figure}

By analyzing the behaviour of the eigenvalues of the inertia tensor, we have 
characterized the increasing anisotropy of a stretched chain.
A useful global index to quantify this anisotropy is provided by the asphericity \cite{rudnickart,rudnicklibro}
\begin{equation}\label{eqn:asphericity_tensor}
\mathcal{A}\equiv \frac{\sum_{s}\left<\left(\lambda_{s}-\bar{\lambda}\right)^{2}\right>}{6\left<\bar{\lambda}^2\right>}=
\frac{3\left<\mathrm{Tr}\,\left(\mathcal{T}^2\right)\right>}{2\left<\left(\mathrm{Tr}\,\mathcal{T}\right)^2\right>}-\frac{1}{2} \,,
\end{equation}
where $\bar{\lambda}=\sum_{s}\lambda_{s}/3$ is the arithmetic mean of the eigenvalues. For a perfectly-symmetric 
distribution, all the eigenvalues have equal magnitude $\lambda_{s}=\bar{\lambda}$,
so that $\mathcal{A}=0$. In the opposite limit of a rod-like chain,
one eigenvalue dominates over the others ($\lambda_1\gg \lambda_2,\lambda_3$) and as a result $\mathcal{A}=1$.
The asphericity of an unperturbed FJC can be computed analytically, and it has 
been shown that, at the leading term, $\mathcal{A}=10/19\simeq 0.53$, independently of $N$ \cite{rudnickart}.
According to Eq. (\ref{eqn:asphericity_tensor}), in order to compute 
$\mathcal{A}$ we need to calculate the averages $\left<\mathrm{Tr}\,\left(\mathcal{T}^2\right)\right>$
and $\left<\left(\mathrm{Tr}\,\mathcal{T}\right)^2\right>$, which
can be written explicitly as
a combination of quadratic terms of $\mathcal{T}$ (see section S4). Such terms can always be decomposed as sums
of end-to-end distances of independent subportions of the chain, whose average can be computed by means of 
our results for $\left<\bm{R_e}^2\right>$ derived above. Proceeding in this way, we could compute analytically 
the asphericity of a stretched FJC and find
\begin{equation}
 \mathcal{A}=1-\frac{72ag+36g^2+24Ng\mathcal{L}^2}{36a^2+80ag+112g^2+4\mathcal{L}^2(11a+10g)N+5N^2\mathcal{L}^4}
 \label{eqn:asphericity_stretch}
\end{equation}
which is shown in Fig.\ref{fig:asph} as a function 
of $\gamma$ for different values of $N$ and is in perfect agreement with the average values of asphericity
from MC simulations.
The limiting behaviors of Eq. (\ref{eqn:asphericity_stretch}) are very instructive.
For $N\mathcal{L}^2\gg 1$, the contributions involving the size of the polymer  
dominate the asphericity, and 
$\mathcal{A}\simeq 1-24/5N\gamma\mathcal{L}$.
A Taylor expansion around $\gamma=0$ gives instead $\mathcal{A}\simeq 10/19 +25N\gamma^2/1083$.
Since at low forces $N\gamma\mathcal{L}\sim N\gamma^2$, we can conclude that the 
asphericity depends on $N$ and $\gamma$ by means of the combination
$N\gamma\mathcal{L}$ in both the limiting cases, which suggests this to be the case, at 
the leading order, in the whole range of forces. As we show in the inset of Fig.\ref{fig:asph},
the data nicely collapse onto the same curve if $\mathcal{A}$ is plotted as a function of $N\gamma\mathcal{L}$.

\begin{figure}[!h]
 \includegraphics[scale=0.21]{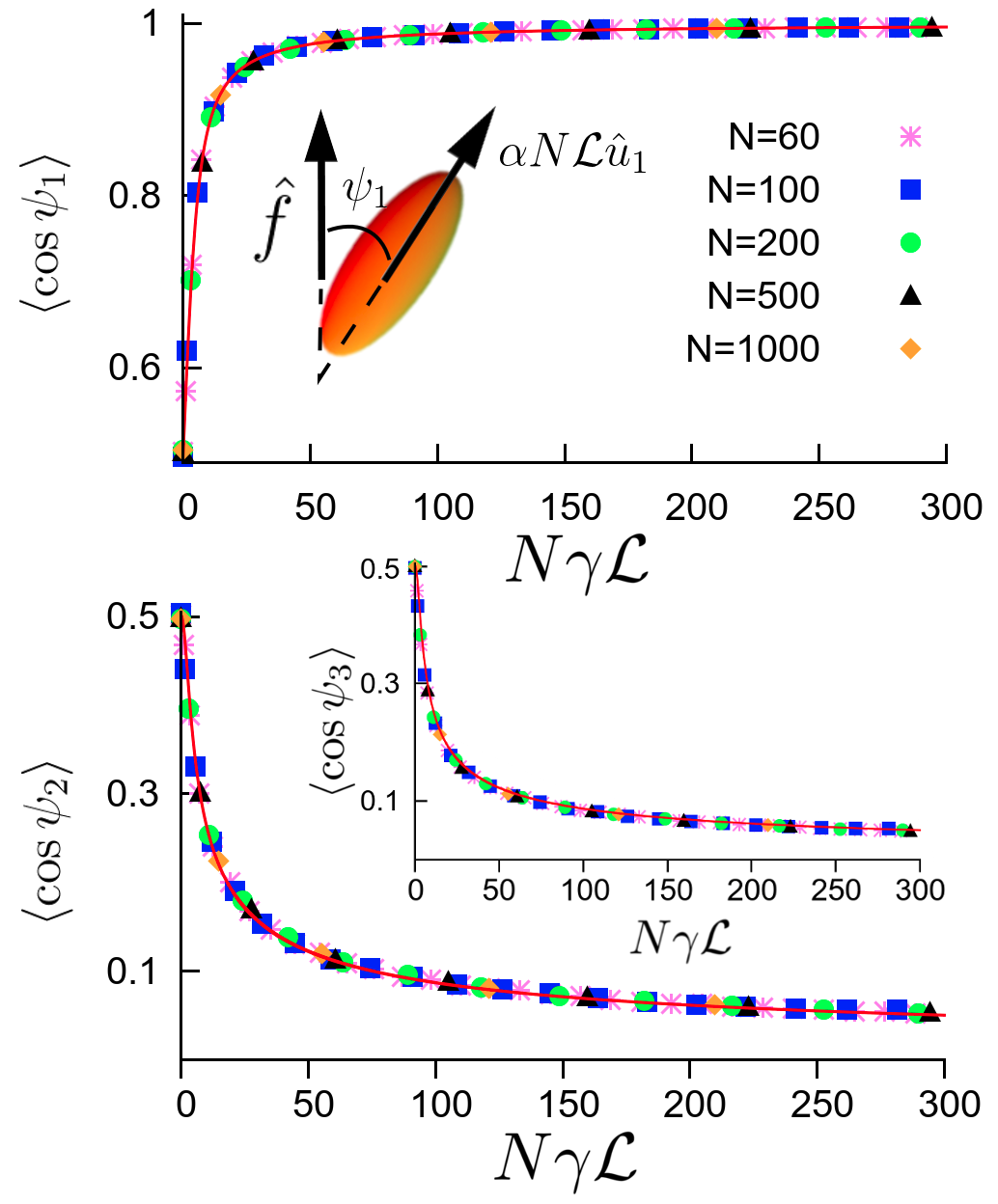}
 \caption{\label{fig:angles_coll} Top panel: collapse of the average value of $\cos\psi_1$ (compare Fig.\ref{fig:angles})
 when plotted as a function of $N\gamma\mathcal{L}$. In the inset we show a schematic picture of the dipole analogy. Bottom panel: 
 collapse of $\left<\cos\psi_2\right>$ and $\left<\cos\psi_3\right>$. The three continuous curves 
 are the predictions obtained from the electric-dipole analogy.}
\end{figure}
Also the orientation of the ellipsoid depends, to the leading order, on the same combination of chain length
and force:
the average cosines $\left<\cos\psi_s\right>$ (compare Fig.\ref{fig:angles} and Fig. S1)
collapse onto a universal curve when plotted as a function of $N\gamma\mathcal{L}$ (Fig.\ref{fig:angles_coll}).
The alignment of the ellipsoid to the external force is analogous to 
the behavior of an electric dipole in the presence of an external field \cite{Neumann2},
although here a $180^{\circ}$ rotation results 
in the same physical state. In this case 
the dipole moment is epitomized by the elongation of the chain, and thus 
we can interpret the factor $N\mathcal{L}$ as being proportional to the polarization response of 
the dipole: larger values of $N$ result into a more responsive chain, although
at large forces the dipole moment saturates to an asymptotic value. In other words, 
we assume the dipole to have a moment equal to $\alpha N\mathcal{L}$, where $\alpha$ 
is a proportionality constant, and to be directed along the main axis of the ellipsoid, as sketched in Fig.\ref{fig:angles_coll}.
Within this assumption, the interaction 
energy is equal to $\alpha N\gamma\mathcal{L}\cos\psi_1$, and the average cosines are equal to 
(see section S5)
\begin{equation}
 \left<\cos\psi_1\right>=
\frac{1}{1-e^{-\alpha N\gamma\mathcal{L}}}-\frac{1}{\alpha N\gamma\mathcal{L}}
\label{eqn:cospsi1_fit}
\end{equation}
and 
\begin{equation}
\left<\cos\psi_2\right>=\left<\cos\psi_3\right>=
\frac{I_1(\alpha N\gamma\mathcal{L})}{\sinh(\alpha N\gamma\mathcal{L})}\,, 
\label{eqn:cospsi2_fit}
\end{equation}
where $I_1$ is the modified Bessel function of the first kind.

The constant $\alpha$ can be determined by considering the large-force behavior. Indeed, from Eq. (\ref{eqn:cospsi1_fit})
we can estimate the average sinus of $\psi_1$ for $\gamma\gg 1$ as
$\left<\sin\psi_1\right>\simeq\sqrt{1-\left<\cos\psi_1\right>^2}\simeq \sqrt{2/\alpha N\gamma\mathcal{L}}$.
Moreover, remembering that $\lambda_1$ is proportional to the square of the best-fitting ellipsoid and that 
its $xy$ projection is given by $c_1 Nb^2g$, by construction we also have 
$\left<\sin\psi_1\right>=\sqrt{c_1 Nb^2g/\left<\lambda_1\right>}\simeq\sqrt{12c_1 Nb^2g/N^2b^2\mathcal{L}^2}=
\sqrt{12c_1/N\gamma\mathcal{L}}$. As a result, we thus find that $\alpha=1/6c_1\simeq 0.83$ which, by means of 
Eqs. (\ref{eqn:cospsi1_fit}) and (\ref{eqn:cospsi2_fit}), leads to the continuous curves showed in Fig.\ref{fig:angles_coll}. 
The remarkable agreement between our ansatz and the simulations shows that 
the dipole analogue can capture even quantitatively the behavior of the ellipsoid in the 
whole range of forces.

In conclusion, starting from the exact distribution of the end-to-end vector, in this work we have characterized in detail the properties 
of a stretched FJC. Our results show that both the shape 
($\left<\lambda_s\right>,\mathcal{A}$) and the orientation ($\left<\cos\psi_s\right>$) of the polymer 
are dominated by finite-size effects. In the case of infinite $N$, any non-zero value of the force 
would result in a rod-like chain 
aligned with the external force, in line with the linear-response ``entropic spring'' result,
according to which for small $\gamma$ the 
relative elongation of the polymer along the direction of the force 
follows a Hooke-like formula $\bm{f}=3k_BT\bm{R_e}/Nb^2$ \cite{rubinstein},
which in the limit of infinite chain length corresponds to an infinitely soft polymer.
Here, we have also quantitatively addressed 
the corrections introduced by finite values of $N$, showing that 
at the leading order both shape and orientation depend on $N$ and $\gamma$ only through 
their combination $N\gamma\mathcal{L}$ and providing analytical formulas for them.
Moreover, we have shown that the transverse section of the ellipsoid shrinks isotropically according to 
novel universal shape features.
Though derived for a FJC, our results for small forces can be directly applied to a wide variety of models,
such as \textit{e.g.} the Wormlike Chain in the case of double-stranded DNA \cite{marko},
provided that the contour length of the chain is much larger than its Kuhn length \cite{rubinstein} but still short
enough so that 
excluded-volume effects can be neglected \cite{strick}.

The authors thank the Swiss National Science Foundation for support under the grants 513469 (P.D.L.R. and A.S.S.)
and 200021-138073 (P.D.L.R. and S.A.). 
\bibliography{quantumbiblio}
\newpage
\vspace*{\fill}
\begin{center}
{\fontsize{18}{21.6} \selectfont \bf Supplemental Material for\\ 
 \vspace{0.4cm}
 The Shape of a Stretched Polymer\\}
  \vspace{0.2cm}
 \fontsize{14}{16.8} \selectfont
 Alberto S. Sassi, Salvatore Assenza \& Paolo De Los Rios
\end{center}
\vspace*{\fill}
\section*{S1 - Exact Computation of $\bm{\left<R_g^2\right>}$}
In the present section, we evaluate the mean squared radius of gyration, extending a method that has been used, 
for example by Rubinstein and Colby \cite{rubinstein}, for the calculation 
of $\left<R_g^2\right>$ of an unperturbed Freely-Jointed Chain.

The radius of gyration can be written as
\begin{equation}
\left<R_g^2\right>=\left<\frac{1}{2(N+1)^2}\sum_{i,j}({\bf R}_i-{\bf R}_j)^2\right> =
\frac{1}{(N+1)^2}\sum_{j>i}\left<({\bf R}_i-{\bf R}_j)^2\right>\ .
\end{equation}
The term inside the two brackets is the end to end vector of the subchain delimited by monomer in position $i$ and $j$.
Implementing the exact formula for $\left<\bm{R_e}^2\right>$ reported in the main text, we easily find 
\begin{equation}
\left<R_g^2\right>= 
\frac{1}{(N+1)^2}\sum_{i=0}^N\sum_{j=i}^N \left[(j-i)b^2\left(1-\mathcal{L}^2\right)+(j-i)^2b^2\mathcal{L}^2\right]\simeq
\frac{1}{6}Nb^2+\frac{1}{12}N^2b^2\mathcal{L}^2\,,
\label{eqn:SI_Rg}
\end{equation}
where we have neglected all the terms which are sublinear in $N$.

The contribution involving only the $x$ and $y$ coordinates can be explicitly computed in a similar fashion. Denoting 
the position vector of the $i$-th monomer as ${\bf R}_i\equiv(X_i,Y_i,Z_i)$
\begin{equation}
 \frac{1}{(N+1)^2}\sum_{j>i}\left<\left(X_j-X_i\right)^2\right>+\left<\left(Y_j-Y_i\right)^2\right>=
 \frac{2}{(N+1)^2}\sum_{j>i}\left<\left(X_j-X_i\right)^2\right>\,,
\end{equation}
where we substituted $\left<\left(Y_j-Y_i\right)^2\right>=\left<\left(X_j-X_i\right)^2\right>$ because
of the cylindrical symmetry of the system.
By means of the exact distribution of the end-to-end vector reported in the main text, we finally find
\begin{equation}
 \frac{2}{(N+1)^2} \sum_{j>i}\left<\left(X_j-X_i\right)^2\right>=
 \frac{2}{(N+1)^2}\sum_{i=0}^N\sum_{j=i}^N (j-i)b^2g\simeq\frac{1}{3}Nb^2g\,.
 \label{eqn:SI_Rgperp}
\end{equation}
The contribution involving the $z$ coordinate can be computed by following the same strategy, yielding as 
a final result
\begin{equation}
 \frac{1}{(N+1)^2} \sum_{j>i}\left<\left(Z_j-Z_i\right)^2\right>\simeq
 \frac{1}{6}Nb^2a+\frac{1}{6}Nb^2\mathcal{L}^2+\frac{1}{12}N^2b^2\mathcal{L}^2=
 \frac{1}{6}Nb^2(1-2g)+\frac{1}{12}N^2b^2\mathcal{L}^2\,.
 \label{eqn:SI_Rgz}
\end{equation}
Naturally, summing Eq. (\ref{eqn:SI_Rgperp}) and Eq. (\ref{eqn:SI_Rgz}) one retrieves the formula for the 
radius of gyration reported in Eq. (\ref{eqn:SI_Rg}).

\section*{S2 - Details of the Monte Carlo simulations}
The main features of the shape of a stretched FJC 
were investigated by means of both analytical computations and MC simulations.
The latter were performed extracting the orientation of each tangent vector 
directly from the distribution $p(\bm{r})$, which we report here for convenience:
\begin{equation}
p(\bm{r})=\frac{\beta fb}{4\pi b^2\sinh(\beta fb)}\exp(\beta\bm{f}\cdot\bm{r})\delta\left(\left|\bm{r}\right|-b\right)\,.
\label{eqn:prob_single_step}
\end{equation}
More in detail, thanks to the cylindric symmetry of the problem, 
the azimuthal angle $\varphi$ could be simply extracted uniformly in 
the range $[0,2\pi]$. As for the polar angle $\vartheta$, a little workaround 
was needed in order to map its distribution onto uniform sampling. 
Let $x$ be a random number uniformly distributed in the range $[0,1]$.
By construction, the infinitesimal probability to find a number in the 
interval $[x,x+dx]$ is simply given by $dx$. As for the angle $\vartheta$,
by considering only the polar contribution to equation (\ref{eqn:prob_single_step})
and remembering that $\gamma\equiv\beta fb$,
we easily find for the distribution of its cosine
\begin{equation}
 \tilde{p}(\cos\vartheta)=\frac{\gamma}{2\sinh\gamma}e^{\gamma\cos\vartheta}\,.
\end{equation}
Since the mapping has to preserve the infinitesimal probability of corresponding 
values of $\theta$ and $x$, the following condition has to be satisfied:
\begin{equation}
\tilde{p}(\cos\vartheta)d\cos\vartheta=dx\,. 
\end{equation}
Integrating both sides, we thus get
\begin{equation}
 \frac{1}{2\sinh \gamma}\left[e^{\gamma\cos\vartheta}-e^{\gamma}\right]=x\,,
\end{equation}
where the integration constant was fixed by imposing $\theta(x=0)=0$.
Therefore, for each step the polar angle $\vartheta$ was computed by inverting 
\begin{equation}
 \cos\vartheta=\frac{\log(e^\gamma-2x\sinh \gamma)}{\gamma}\,.
\end{equation}
For each value of $N$ and $\gamma$, the mean values of the several
quantities considered in the main text
are obtained by averaging $10^4$
different realizations. Statistical error is estimated by normalizing the standard deviation of 
the results and, when not shown, is always smaller than the size of symbols in the figures.

\section*{S3 - Analysis of subleading terms in $\bf{\lambda_1}$}
In the present section we focus on the various terms contributing to $\lambda_1$ in the large-force regime.
In section S1 we computed the contributions to the radius of gyration coming from
the $xy$ projection (Eq. (\ref{eqn:SI_Rgperp})) and the $z$ component (Eq. (\ref{eqn:SI_Rgz}))
of the FJC.
For $\gamma\gg1$, $\lambda_1$ is expected to give the leading contribution to 
the $z$ component as well as to play a 
significant role in quantitatively determining the $xy$ projection.
Therefore, we predict the following functional form:
\begin{equation}
 \left<\lambda_1\right>=\frac{1}{6}Nb^2(1-2g)+\frac{1}{12}N^2b^2\mathcal{L}^2+c_1Nb^2g\,.
 \label{eqn:SI_lambda1}
\end{equation}
Starting from the MC data reported in Fig. 3 top in the main text, we thus considered the combination
\begin{equation}
 \lambda_1^{\mbox{\tiny sub}}\equiv \left<\lambda_1\right>-\frac{1}{6}Nb^2(1-2g)-\frac{1}{12}N^2b^2\mathcal{L}^2\,.
\end{equation}
According to Eq. (\ref{eqn:SI_lambda1}), the following formula should thus hold
\begin{equation}
 \frac{\lambda_1^{\mbox{\tiny sub}}}{Nb^2}=c_1g\,.
 \label{eqn:SI_lambda1_sub}
\end{equation}
As we show in Fig.\ref{fig:lambda1_sub}, the MC data collapse onto a universal 
curve when $\lambda_1^{\mbox{\tiny sub}}$ is normalized by the chain size $N$. Moreover, 
by tuning the numeric constant $c_1$ they are well described by the function $g$, as expected from
Eq. (\ref{eqn:SI_lambda1_sub}). The optimum value of the constant is $c_1=0.204\pm 0.001$, in perfect
agreement with the result found in the main text starting from the fits of $\lambda_2$ and $\lambda_3$.

\section*{S4 - Exact Computation of Asphericity}
In the case of the asphericity the idea is the same as for the radius of gyration, but a larger number of terms must be evaluated. 
We first rewrite the formula in the following way: 
\begin{equation}\label{defas}
\left<A\right> \ = \ \frac{3\left<Tr\left[\mathcal{T}^2\right]\right>}{2\left<\left(Tr\left[\mathcal{T}\right]\right)^2\right>}-\frac{1}{2} \ \equiv
\ \frac{3\sum_{\alpha=1}^{3}\sum_{\beta=1}^{3}\left<\mathcal{T}_{\alpha\beta}\mathcal{T}_{\beta\alpha}\right>}{2\sum_{\alpha=1}^{3}\sum_{\beta=1}^{3}\left<\mathcal{T}_{\alpha\alpha}\mathcal{T}_{\beta\beta}\right>}-\frac{1}{2} \ ,
\end{equation}  
where we are using Greek letters as labels for spatial coordinates ($x,y,z$).
We have to find the values of two terms, that we write more explicitly:
\begin{equation}
\left<\mathcal{T}_{\alpha\beta}\mathcal{T}_{\beta\alpha}\right> = \frac{1}{N^4}\sum_{i=0}^{N}\sum_{j=i}^{N}\sum_{k=0}^{N}\sum_{l=k}^{N}
\left<\left(\alpha_{j}-\alpha_{i}\right)\left(\beta_{j}-\beta_{i}\right)
\left(\alpha_{l}-\alpha_{k}\right)\left(\beta_{l}-\beta_{k}\right)\right> \ ,
\end{equation}
\begin{equation}
\left<\mathcal{T}_{\alpha\alpha}\mathcal{T}_{\beta\beta}\right> \ = \ \frac{1}{N^4}\sum_{i=0}^{N}\sum_{j=i}^{N}\sum_{k=0}^{N}\sum_{l=k}^{N}
\left<\left(\alpha_{j}-\alpha_{i}\right)^{2}\left(\beta_{l}-\beta_{k}\right)^{2}\right> \ .
\end{equation}\vspace{0.4cm}\\
Since the tensor is symmetric, only six terms are independent: $\mathcal{T}_{xx}^2$, $\mathcal{T}_{xy}^2$, $\mathcal{T}_{xz}^2$, $\mathcal{T}_{zz}^2$, $\mathcal{T}_{xx}\mathcal{T}_{yy}$,
$\mathcal{T}_{xx}\mathcal{T}_{zz}$.
The most complicated are the ones with equal indices, $\mathcal{T}_{xx}^2$ and $\mathcal{T}_{zz}^2$.
We will calculate $\mathcal{T}_{zz}^2$. All the other terms can be evaluated in a similar way. We must treat separately three different cases:\vspace{0.4cm}\\
\fbox{$I)$ \  $i<j<k<l$}\vspace{0.8cm}\\
\begin{figure}[H]
\centering
\includegraphics[scale=0.15]{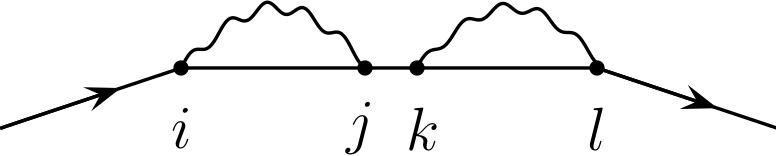}
\end{figure}
In this case there is no overlap between the intervals $\left(i,j\right)$ and $\left(k,l\right)$ and the mean of a product becomes the product of means:\vspace{0.4cm}\\
\begin{align}
[I]  &= \  2\frac{1}{N^4}\sum_{i=0}^{N}\sum_{j=i}^{N}\sum_{k=j}^{N}\sum_{l=k}^{N}\left<\left(z_{j}-z_{i}\right)^{2}\left(z_{l}-z_{k}\right)^{2}\right>  = \\  
&= 2\frac{1}{N^4}\sum_{i=0}^{N}\sum_{j=i}^{N}\sum_{k=j}^{N}\sum_{l=k}^{N}\left<\left(z_{j}-z_{i}\right)^{2}\right>\left<\left(z_{l}-z_{k}\right)^{2}\right> \ = \\ 
&= 2\frac{1}{N^4}\sum_{i=0}^{N}\sum_{j=i}^{N}\left[(j-i)b^2a+{\cal L}^2(j-i)^2b^2\right]\sum_{k=j}^{N}\sum_{l=k}^{N}\left[(l-k)b^2a+{\cal L}^2(l-k)^2b^2\right] \ ,
\end{align}
where $a=1-2g-{\cal L}^2$ and we have used the result previously obtained for the $z$ component of the end to end vector. 
\vspace{1 cm}\\
\fbox{$II)$ \  $i<k<j<l$}
\begin{figure}[H]
\centering
\includegraphics[scale=0.14]{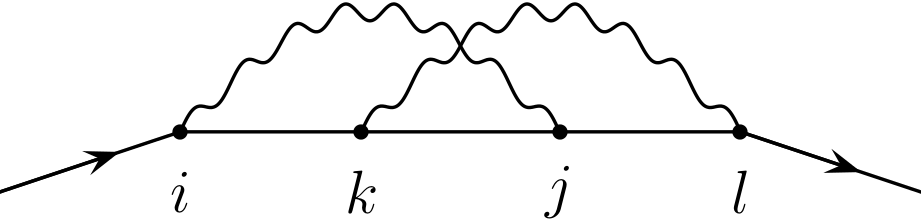}
\end{figure}
\begin{equation}
[II] \ = \  2\frac{1}{N^4}\sum_{i=0}^{N}\sum_{k=i}^{N}\sum_{j=k}^{N}\sum_{l=j}^{N}\left<\left(z_{j}-z_{i}\right)^{2}\left(z_{l}-z_{k}\right)^{2}\right> \ .
\end{equation}
Because of the overlap the average cannot be split {\em directly}. However, this problem can be solved with a trick: 
\begin{equation*}
\begin{split}
& \left<\left(z_{j}-z_{i}\right)^{2}\left(z_{l}-z_{k}\right)^{2}\right> \ = \ \left<\left(\left(z_{j}-z_{k}+z_{k}-z_{i}\right)
\left(z_{l}-z_{j}+z_{j}-z_{k}\right)\right)^{2}\right> \ = \\
& \left<\left(\left(z_{j}-z_{k}\right)\left(z_{l}-z_{j}\right)+\left(z_{k}-z_{i}\right)\left(z_{l}-z_{j}\right)+
\left(z_{j}-z_{k}\right)\left(z_{k}-z_{i}\right)+\left(z_{j}-z_{k}\right)^2\right)^{2}\right> \ = \\
& \left<\left(z_{j}-z_{k}\right)^{2}\left(z_{l}-z_{j}\right)^{2}\right>+\left<\left(z_{k}-z_{i}\right)^{2}\left(z_{l}-z_{j}\right)^{2}\right>+
\left<\left(z_{j}-z_{k}\right)^{2}\left(z_{k}-z_{i}\right)^{2}\right>+\left<\left(z_{j}-z_{k}\right)^{4}\right> \\
& +2\left<\left(z_{j}-z_{k}\right)\left(z_{k}-z_{i}\right)\left(z_{l}-z_{j}\right)^{2}\right>+
2\left<\left(z_{j}-z_{k}\right)\left(z_{l}-z_{j}\right)\left(z_{k}-z_{i}\right)^{2}\right>  \\
& +4\left<\left(z_{l}-z_{j}\right)\left(z_{k}-z_{i}\right)\left(z_{j}-z_{k}\right)^{2}\right>+  2\left<\left(z_{l}-z_{j}\right)\left(z_{j}-z_{k}\right)^{3}\right>+2\left<\left(z_{k}-z_{i}\right)\left(z_{j}-z_{k}\right)^{3}\right> \ .
\end{split}
\end{equation*}
The final expression, even if it is much longer, has a great advantage: now in any product the round brackets are uncorrelated with respect to each other, and we can split the averages:
\begin{equation*}
<\left( \ \right)...\left( \ \right)> \ = \ <\left( \ \right)>....<\left( \ \right)> \ .
\end{equation*}
In this way, each term in the previous equation can be computed as above.
\vspace{1 cm}\\
\fbox{$III)$ \  $i<k<l<j$}
\begin{figure}[H]
\centering
\includegraphics[scale=0.16]{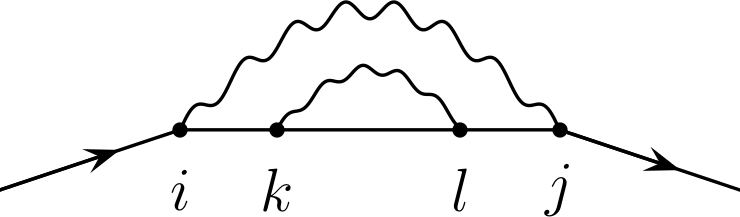}
\end{figure}
\begin{equation}
[III] \ = \  2\frac{1}{N^4}\sum_{i=0}^{N}\sum_{k=i}^{N}\sum_{l=k}^{N}\sum_{j=l}^{N}<\left(z_{j}-z_{i}\right)^{2}\left(z_{l}-z_{k}\right)^{2}> \ .
\end{equation}
Since the calculation is analogous to the previous one, we will just show how it is possible to decorrelate each term:
\begin{equation}
<\left(z_{j}-z_{i}\right)^{2}\left(z_{l}-z_{k}\right)^{2}> \ = \ <\left(\left(z_{j}-z_{l}+z_{l}-z_{k}+z_{k}-z_{i}\right)
\left(z_{l}-z_{k}\right)\right)^{2}> \ ,
\end{equation}
and then the usual calculation is made. 
For evaluating the mean values we need the following expressions:
\begin{equation}
\begin{split}
& {\sigma^2}_z = Nb^{2}(1-2g-\mathcal{L}^2) \\
& \mu_z = Nb\mathcal{L} \\
& {\sigma^2}_{x,y} = Nb^{2}g\\
& \mu_{x,y} = 0\\
& \left<\alpha\right> \ = \mu_{\alpha}\\
& \left<\alpha^2\right> \  = \mu_{\alpha}^2+\sigma_{\alpha}^{2}\\
& \left<\alpha^3\right> \  = \mu_{\alpha}^3+3\mu_{\alpha}\sigma_{\alpha}^{2}\\
& \left<\alpha^4\right> \  =  \mu_{\alpha}^4+3\sigma_{\alpha}^4+6\mu_{\alpha}^{2}\sigma_{\alpha}^{2} \ .
\end{split}
\end{equation}
where $\alpha$ indicates a space coordinate. Keeping only the leading terms, the result is
\begin{equation}
\left<\mathcal{T}_{zz}^2\right> \ = \ [I] \ + \ [II] \ + \  [III] \ = \ \frac{N^2b^4}{720}
\left(5\mathcal{L}^{4}N^{2}+44\mathcal{L}^{2}Na+36a^2\right) \ ,
\end{equation}
The six moments are
\begin{equation}
\begin{split}
& \left<\mathcal{T}_{xx}^2\right> \ = \ \frac{g^{2}N^{2}b^4}{20}\\
& \left<\mathcal{T}_{xy}^2\right> \ = \ \frac{g^{2}N^{2}b^4}{90}\\
& \left<\mathcal{T}_{xx}\mathcal{T}_{yy}\right> \ = \ \frac{g^{2}N^{2}b^4}{36}\\
& \left<\mathcal{T}_{zz}^2\right> \ = \ \frac{N^2b^4}{720}\left(5\mathcal{L}^{4}N^{2}+44\mathcal{L}^{2}Na+36a^2\right)\\
& \left<\mathcal{T}_{xz}^2\right> \ = \ \frac{gN^2b^4}{360}\left(4a+3\mathcal{L}^{2}N\right)\\
& \left<\mathcal{T}_{xx}\mathcal{T}_{zz}\right> \ = \ \frac{gN^2b^4}{72}\left(2a+\mathcal{L}^{2}N\right) \ .
\end{split}
\end{equation}
Substituting in equation (S4), we finally find the mean asphericity
\begin{equation}\label{asph2}
\left<A\right> = 1 - \frac{72ag+36g^2+24Ng\mathcal{L}^2}{36a^2+80ag+112g^2+4\mathcal{L}^{2}\left(11a+10g\right)N+5N^{2}\mathcal{L}^4} \ .
\end{equation}
which is the formula reported in the main text.

\section*{S5 - Computation of ellipsoid orientation within the dipole analogy}
As explained in the main text, we captured the orientational behavior of the enveloping ellipsoid of a 
stretched FJC by exploiting the strong resemblance of this system to an electric dipole in the presence
of an external field. Starting from the interaction energy $\alpha N\gamma\mathcal{L}\cos\psi_1$, 
the average cosine of the main axis of the ellipsoid can be straightforwardly computed:
\begin{equation}
 \left<\cos\psi_1\right>=\frac{\int_0^{\pi/2}\sin\psi_1\cos\psi_1 e^{\alpha N\gamma\mathcal{L}\cos\psi_1}d\psi_1}
 {\int_0^{\pi/2} \sin\psi_1e^{\alpha N\gamma\mathcal{L}\cos\psi_1}d\psi_1}=
\frac{1}{1-e^{-\alpha N\gamma\mathcal{L}}}-\frac{1}{\alpha N\gamma\mathcal{L}}\,,
\end{equation}
which is Eq. (4) in the main text. We note that the upper bound of the integrals in the previous formula 
is $\pi/2$ because of the symmetry of the system with respect to a rotation by $\pi$. 

In contrast, the computation of $\left<\cos\psi_2\right>$ (which, due to the cylindrical 
symmetry of the dipole analogy is equal to $\left<\cos\psi_3\right>$)
is more cumbersome, and we report its explicit derivation in what follows. The following formulas
will be needed for our computation:
\begin{equation}
 \int_0^{2\pi} \left(\sin \theta\right)^k d\theta=
 \begin{cases}
  0 & \mbox{if $k$ odd}\\
  2\pi \frac{\left(k-1\right)!!}{k!!} & \mbox{if $k$ even}
 \end{cases}\,,
 \label{eqn:SI_int1}
\end{equation}
\begin{equation}
 \int_0^{\pi/2}\left(\sin\psi\right)^{2k+1}d\psi=\frac{\left(2k\right)!!}{\left(2k+1\right)!!}\,,
 \label{eqn:SI_int2}
\end{equation}
\begin{equation}
 I_1(x)=\sum_{k=0}^{\infty}\frac{1}{k!(k+1)!}\left(\frac{x}{2}\right)^{2k+1}\,,
 \label{eqn:SI_int3}
\end{equation}
\begin{equation}
 \sinh x=\sum_{k=0}^{\infty}\frac{x^{2k+1}}{(2k+1)!}\,,
 \label{eqn:SI_int4}
\end{equation}
where in Eqs. (\ref{eqn:SI_int1}) and (\ref{eqn:SI_int2}) it is intended 
that the double factorials are equal to $1$ for $k=0$, and
in Eq. (\ref{eqn:SI_int3}) $I_1(x)$ is the modified Bessel function of the first kind.
For a given value of $\psi_2$, one has to consider 
for the eigenvector $\hat{u}_1$ corresponding 
to $\lambda_1$ all the possible orientations lying on 
the plane perpendicular to the versor $\hat{u}_2$ identified by $\psi_2$. According to the relative 
orientation of $\hat{u}_1$ and the external force, each specific orientation results into 
a different interaction energy, \textit{i.e.} a different
Boltzmann weight. Quantitatively, let us consider a reference frame where the $z$ axis
is oriented along the external force and the $x$ axis is complanar to both $\hat{z}$ and $\hat{u}_2$.
In this reference frame, one has $\hat{u}_2\equiv\left(\sin\psi_2,0,\cos\psi_2\right)$. From here, it is 
easy to show that a unitary vector lying on the plane perpendicular to $\hat{u}_2$ can be written as 
$\left(-\cos\psi_2\sin\theta,\cos\theta,\sin\psi_2\sin\theta\right)$, where $\theta\in[0,2\pi]$.
Therefore, the orientation of $u_1$ with respect to the force is given by $\cos\psi_1=\sin\psi_2\sin\theta$, and 
the corresponding adimensionalized interaction energy between the dipole and the external field is 
$\alpha N\gamma\mathcal{L}\cos\psi_1=\alpha N\gamma\mathcal{L}\sin\psi_2\sin\theta$.
The total Boltzmann weight corresponding to the given choice of $\psi_2$ is obtained by integrating 
the Boltzmann weights relative to this energy over all the possible values of $\theta$. Thus
\begin{equation}
 \left<\cos\psi_2\right>=\frac{\int_0^{\pi/2}\sin\psi_2\cos\psi_2
 \left(\int_0^{2\pi} e^{\alpha N\gamma\mathcal{L}\sin\psi_2\sin\theta}d\theta\right)d\psi_2}
 {\int_0^{\pi/2}\sin\psi_2
 \left(\int_0^{2\pi} e^{\alpha N\gamma\mathcal{L}\sin\psi_2\sin\theta}d\theta\right)d\psi_2}\,.
 \label{eqn:formula_init_psi2}
\end{equation}
We first address the computation of the Boltzmann weight. By Taylor-expanding the exponential, we have
\begin{equation}
 \int_0^{2\pi} e^{\alpha N\gamma\mathcal{L}\sin\psi_2\sin\theta}d\theta=
 \sum_{k=0}^{\infty}\frac{\left(\alpha N\gamma\mathcal{L}\sin\psi_2\right)^k}{k!}
 \int_0^{2\pi}\left(\sin\theta\right)^kd\theta\,.
\end{equation}
Making use of Eq.(\ref{eqn:SI_int1}) and performing the change of dummy variable $k=2m$ in the 
non-vanishing terms of the series, we find
\begin{equation}
 \int_0^{2\pi} e^{\alpha N\gamma\mathcal{L}\sin\psi_2\sin\theta}d\theta=
 2\pi\sum_{m=0}^{\infty}\frac{\left(\alpha N\gamma\mathcal{L}\right)^{2m}}{\left(2m\right)!}
 \frac{\left(2m-1\right)!!}{\left(2m\right)!!}\left(\sin\psi_2\right)^{2m}\,.
\end{equation}
The numerator in Eq.(\ref{eqn:formula_init_psi2}) thus becomes
\begin{equation*}
 \int_0^{\pi/2}\sin\psi_2\cos\psi_2
 \left(\int_0^{2\pi} e^{\alpha N\gamma\mathcal{L}\sin\psi_2\sin\theta}d\theta\right)d\psi_2=
\end{equation*}
\begin{equation*}
=2\pi\sum_{m=0}^{\infty}\frac{\left(\alpha N\gamma\mathcal{L}\right)^{2m}}{\left(2m\right)!}
 \frac{\left(2m-1\right)!!}{\left(2m\right)!!}
 \int_0^{\pi/2}\cos\psi_2\left(\sin\psi_2\right)^{2m+1}d\psi_2=
\end{equation*}
\begin{equation*}
=2\pi\sum_{m=0}^{\infty}\left(\alpha N\gamma\mathcal{L}\right)^{2m}
 \frac{\left(2m-1\right)!!}{\left(2m\right)!}
 \frac{\left.\left(\sin\psi_2\right)^{2m+2}\right|_{\psi_2=0}^{\pi/2}}{\left(2m+2\right)\left(2m\right)!!}=
\end{equation*}
\begin{equation*}
=2\pi\sum_{m=0}^{\infty}
 \frac{\left(\alpha N\gamma\mathcal{L}\right)^{2m}}{\left(2m+2\right)!!\left(2m\right)!!}\,,
\end{equation*}
where in the last step we exploited the identity $\left(2m\right)!=\left(2m\right)!!\left(2m-1\right)!!$. Moreover, 
since $\left(2m\right)!!=2^mm!$, the previous formula can be rewritten as
\begin{equation}
2\pi\sum_{m=0}^{\infty}
 \frac{\left(\alpha N\gamma\mathcal{L}\right)^{2m}}{\left(2m+2\right)!!\left(2m\right)!!}= 
 \frac{2\pi}{\alpha N\gamma\mathcal{L}}\sum_{m=0}^{\infty}
 \frac{1}{m!\left(m+1\right)!}\left(\frac{\alpha N\gamma\mathcal{L}}{2}\right)^{2m+1}=
 2\pi\frac{I_1(\alpha N\gamma\mathcal{L})}{\alpha N\gamma\mathcal{L}}\,,
 \label{eqn:SI_num}
\end{equation}
where we made use of Eq. (\ref{eqn:SI_int3}). Analogously, the denominator in Eq. (\ref{eqn:formula_init_psi2})
can be explicitely computed as 
\begin{equation*}
 \int_0^{\pi/2}\sin\psi_2
 \left(\int_0^{2\pi} e^{\alpha N\gamma\mathcal{L}\sin\psi_2\sin\theta}d\theta\right)d\psi_2=
\end{equation*}
\begin{equation*}
=2\pi\sum_{m=0}^{\infty}\frac{\left(\alpha N\gamma\mathcal{L}\right)^{2m}}{\left(2m\right)!}
 \frac{\left(2m-1\right)!!}{\left(2m\right)!!}
 \int_0^{\pi/2}\left(\sin\psi_2\right)^{2m+1}d\psi_2=
\end{equation*}
\begin{equation*}
=2\pi\sum_{m=0}^{\infty}\frac{\left(\alpha N\gamma\mathcal{L}\right)^{2m}}{\left(2m\right)!}
 \frac{\left(2m-1\right)!!}{\left(2m\right)!!}
 \frac{\left(2m\right)!!}{\left(2m+1\right)!!}=
\end{equation*}
\begin{equation*}
=\frac{2\pi}{\alpha N\gamma\mathcal{L}}
\sum_{m=0}^{\infty}\frac{\left(\alpha N\gamma\mathcal{L}\right)^{2m+1}}{\left(2m+1\right)!}=
\end{equation*}
\begin{equation}
=2\pi\frac{\sinh(\alpha N\gamma\mathcal{L})}{\alpha N\gamma\mathcal{L}}\,,
\label{eqn:SI_den}
\end{equation}
where in the second step we substituted Eq. (\ref{eqn:SI_int2}), while in the last step we 
made use of the Taylor expansion reported in Eq. (\ref{eqn:SI_int4}). Substituting 
Eq.(\ref{eqn:SI_num}) and Eq.(\ref{eqn:SI_den}) into Eq.(\ref{eqn:formula_init_psi2}), we 
finally obtain
\begin{equation}
 \left<\cos\psi_2\right>=\frac{I_1(\alpha N\gamma\mathcal{L})}{\sinh(\alpha N\gamma\mathcal{L})}\,,
\end{equation}
which is the result reported in Eq. (5) in the main text.
\begin{figure}[H]
\centering
 \includegraphics[scale=0.36]{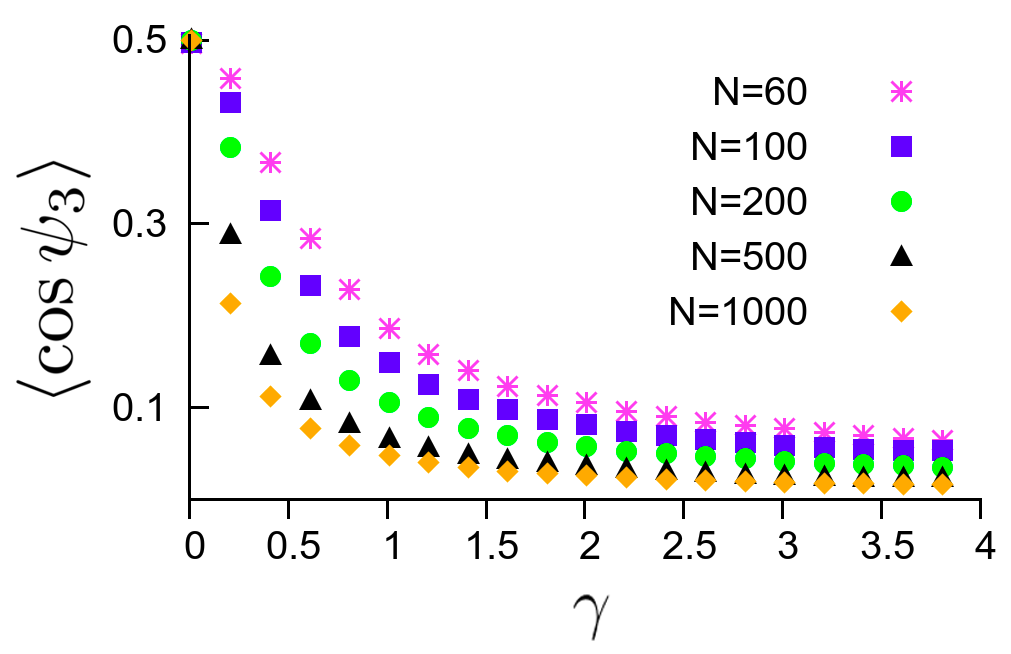}
 \caption{\label{fig:angles_l3} Average cosine of the angle between the external force 
 and the eigenvector corresponding to $\lambda_3$.}
\end{figure}
\begin{figure}[H]
\centering
 \includegraphics[scale=0.36]{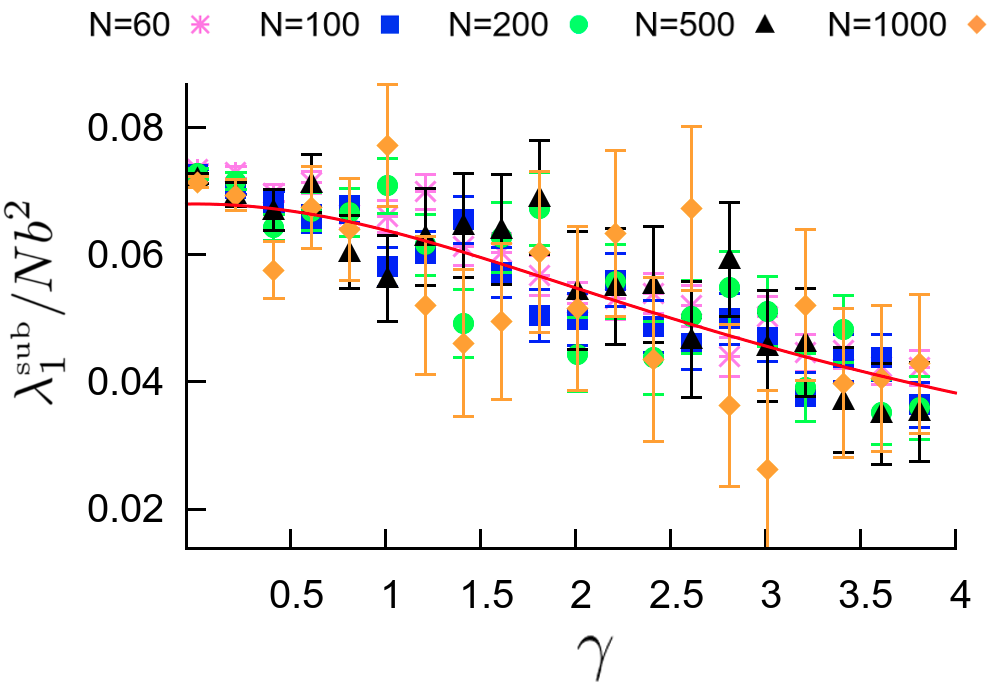}
 \caption{\label{fig:lambda1_sub} 
 Plot of $\lambda_1^{\mbox{\tiny sub}}/Nb^2$ as a function of $\gamma$ for several values of $N$.}
\end{figure}
\end{document}